\documentclass[12pt,a4paper]{article}
\pdfoutput=1
\usepackage{jheppub}
\makeatletter
\def\@fpheader{\relax}
\makeatother

\def\cP{{\cal P}}
\def\cK{{\cal K}}

\title{SYK Model, Chaos and Conserved Charge}
\author{Ritabrata Bhattacharya$^{a}$, Subhroneel Chakrabarti$^{a}$, Dileep P.~Jatkar$^{a}$, Arnab Kundu$^{b}$}
\affiliation{$^a$Harish-Chandra Research Institute(HBNI), Chhatnag
  Road, Jhusi, Allahabad 211019, India.}

\affiliation{$^b$Theory Division, Saha Institute of Nuclear Physics(HBNI), 1/AF Bidhannagar, Kolkata 700064, India.}

\emailAdd{ritabratabhattacharya[at]hri.res.in, subhroneelchak[at]hri.res.in, dileep[at]hri.res.in, arnab.kundu[at]saha.ac.in} 

\abstract{We study the SYK model with complex fermions, in the presence of an all-to-all $q$-body interaction, with a non-vanishing chemical
potential.  We find that, in the large $q$ limit, this model can be
solved exactly and the corresponding Lyapunov exponent can be obtained semi-analytically. The resulting Lyapunov exponent is a sensitive
function of the chemical potential $\mu$.  Even when the coupling $J$, which
corresponds to the disorder averaged values of the all to all fermion
interaction, is large, values of $\mu$ which are exponentially small
compared to $J$ lead to suppression of the Lyapunov exponent.}

\begin{document}

\maketitle
\flushbottom


\section{Introduction} 

Given a quantum dynamical system, {\it e.g.},~a specific Hamiltonian, a
ubiquitous feature is the chaotic property of the same, which
subsequently leads to ergodicity, thermalization and similar universal
and coarse-grained description\cite{strogatz}. For
classical dynamical systems, the measure of chaos is simple: a
response of the classical trajectories with respect to initial
conditions\cite{strogatz}. Quantum mechanically,
although may not be unique, a quantitative measure can be given in
terms of the square of commutators of self-adjoint operators that are
time-separated. Specifically, from the large time behaviour of the
same which typically takes the form of an exponential growth in time,
one can extract the quantum analogue of the Lyapunov exponent, that,
for classical dynamical systems, measures the sensitivity of two
initially nearby (in the space of initial conditions) trajectories
with respect to the corresponding initial conditions, as time evolves
to large values.

The definition of the Lyapunov exponent, and equivalently the
quantitative notion of quantum chaos, is naturally associated with
large time limit of a dynamical system and therefore can be
interpreted as an inherently infra-red (IR) quantity. Within the
purview of quantum field theory (QFT) {\it a la} Wilson, one begins
with an ultra-violet (UV) description of a system, subsequently
integrates out the massive modes and arrives at an effective IR
description. Given a QFT at the UV, the corresponding Lyapunov
exponent can be extracted from the large time behaviour of
out-of-time-ordered (OTO) correlation function\cite{larkin}. The
resulting Lyapunov exponent is a non-trivial function of the
dimensionless couplings that define the UV-theory. In principle, a
renormalization group (RG) flow maps the set of UV couplings to a set
of IR-couplings, and thus the Lyapunov exponent is a different
non-trivial function of the IR-couplings\cite{Shenker:2013pqa}.

In general, given a QFT or a quantum mechanical system, it is
non-trivial to obtain the Lyapunov exponent. Recently, a lot of
progress has been made in the Sachdev-Ye-Kitaev (SYK)
model\cite{SY,Kitaev}, in which the Lyapunov exponent has been
analytically calculated (see {\it e.g.}~\cite{Maldacena:2016hyu}) and
subsequently demonstrated to satisfy the maximal chaos bound, proposed
and argued in \cite{Maldacena:2015waa}. Motivated by this, specially
the saturation of the maximal bound which is thought to be a necessary
condition for a quantum system to have a holographic description,
connections of AdS$_2$/CFT$_1$ have been explored further, beginning
with \cite{Maldacena:2016upp} and followed up by a large volume of
work on the SYK model and its various generalisations, involving
complex fermions, tensor models and higher dimensional
analogs\cite{Giombi:2017dtl,Klebanov:2017nlk,
  Gross:2016kjj,Bonzom:2017pqs,Gu:2016oyy,Garcia-Garcia:2017bkg,Polchinski:2016xgd,
  Klebanov:2016xxf,Fu:2016vas,Witten:2016iux,Nishinaka:2016nxg,
  Gurau:2016lzk,Krishnan:2016bvg,Gurau:2017qna,Turiaci:2017zwd,Narayan:2017qtw,Krishnan:2017txw,
  Berkooz:2016cvq, Jevicki:2016bwu}.  Supersymmetric generalisations
of the SYK model have also been
studied\cite{Fu:2016vas,Murugan:2017eto,Yoon:2017gut}.  The proposed
holographic dual in terms of the Schwarzian action has also been
analysed\cite{Stanford:2017thb}.

Although the precise connection in terms of AdS/CFT remains unclear,
the SYK model is undoubtedly unique in capturing the following
features, all at once: solvable at large $N$, emergence of conformal
invariance in the IR and maximal chaos. Emergence of AdS$_2$, on the
other hand, is rather unique in stringy physics: ranging from the
entropy counting of extremal black hole horizons to the emergent IR
description of a large $N$ strongly coupled gauge theory with
non-vanishing density. Thus, a physical result obtained from the SYK
model is likely to be relevant about the physics of AdS$_2$, viewed in
the appropriate context.

In this article, we explore a simple way to tune chaos in the SYK-type
model, by introducing global conserved charges. We focus on the
SYK-model with complex fermions, that have previously been studied in
{\it e.g.}~\cite{Davison:2016ngz, Bulycheva:2017uqj}, where the
fermions have an all-to-all $q$-body interaction, with Gaussian random
distribution for the coupling strength. The standard SYK model
corresponds to taking $q=4$, however, similar to
\cite{Bulycheva:2017uqj}, we study the limit of $q \to \infty$. This
limit particularly facilitates analytical calculations, where much of
the large $q$ analysis of \cite{Maldacena:2016hyu} can be generalized
in the presence of a non-vanishing global charge, to obtain the
corresponding Lyapunov exponent.  There is another method of getting
tunable Lyapunov exponent by coupling peripheral fermions to the SYK
model \cite{Banerjee:2016ncu}.

In this case, the UV-theory comes equipped with two independent
couplings: $\left\{ \beta J, \beta \mu \right \}$, where
$\beta = T^{-1}$ is the inverse temperature, $J$ measures the
interaction strength after performing a random averaging, and $\mu$ is
the chemical potential corresponding to the global charge, which
introduces a new scale in the problem. In the large $N$, large $q$
limit, with $N \ge q$, the resulting Schwinger-Dyson equation regroups
the UV couplings to an effective IR coupling, such that the RG flow
maps $\left\{ \beta J, \beta \mu \right \} \to \beta \tilde{J}$. The
large $q$ analysis now yields the Lyapunov exponent:
$\lambda_{\rm L} \left( \beta \mu, \beta J \right) \equiv \lambda_{\rm
  L} \left( \beta \tilde{J} \right) $. It turns out that, by tuning
the UV data one can smoothly interpolate between
$\lambda_{\rm L} = 2 \pi T$ to $\lambda_{\rm L} = 0$. We also obtain a
similar result for the complex fermions with a global flavour
symmetry, introduced in \cite{Gross:2016kjj}. The non-invertible map
of the couplings, from UV to IR, emerges at large $q$, even when
sub-leading effects in $(1/q)$ are considered, and may be an artefact
of this limit.\footnote{This is an interesting issue to explore
  further. It may happen that an {\it attractor type} behaviour
  exists, in which the deep IR physics reorganizes itself in terms of
  emerging parameters, irrespective of the value of $q$. }

That the presence of global charges suppress the Lyapunov exponent
and, in fact, can tune it to vanishing values is no
surprise. Conserved charges constrain the phase-space of any dynamical
system, the extreme limit of which are represented by integrable
models. For the latter, no chaotic behaviour is expected. Thus, the
result above interpolates between a chaotic behaviour to a non-chaotic
regime, even with a U$(1)$ charge, as the corresponding chemical
potential is increased. Similar feature upholds for the flavoured
complex fermion models.

In terms of the Schwinger-Dyson equation, at large $N$, exploring the
strong coupling phase of the system is equivalent to taking a deep IR
limit. Keeping the effect of a non-vanishing chemical potential is
similar to working at an intermediate energy-scale. In fact,
Schwinger-Dyson equation comes equipped with a term
$\left( i \omega + \mu \right) $, where $\omega$ is the frequency, and
thus $\mu$ and $\omega$ seem freely tradable. From the UV-perspective,
the interpretation is physically distinct, but for an
intermediate-scale observer, studying $\lambda_{\rm L}$ as a function
of $\beta\mu$ is similar to studying how $\lambda_{\rm L}$ changes
away from the IR-conformal limit.\footnote{We note here that, even in
  the presence of a chemical potential, that defines a scale for the
  system, in the deep IR conformal symmetry is recovered, when
  supplemented by a gauge transformation. See {\it
    e.g.}~\cite{Davison:2016ngz}.} The dependence of $\lambda_{\rm L}$
with $\beta J$ have already been explored in \cite{Maldacena:2016hyu},
in the $q \to \infty$ limit.

This article is divided in the following sections: In section $2$, we briefly introduce the model with complex fermions, obtain the Schwinger-Dyson equation and present the solution in the $q \to \infty$ limit. We subsequently discuss the calculation of the retarded kernel in the next section. Section $4$ is devoted to studying the dependence of the Lyapunov exponent, in details. We comment briefly on flavoured complex fermion model in section $5$. Finally, we conclude with future directions.

\section{SYK model with complex fermions and chemical potential}

\subsection{The SYK model}
\label{sec:syk-model}

We will begin by briefly recalling the SYK model.  The SYK model
describes all-to-all random interactions between $N$ Majorana fermions
in $(0+1)$ dimension involving $q$ fermions at a time.  The Hamiltonian is
given by\cite{Kitaev, Maldacena:2016hyu}
\begin{eqnarray}
H = \left( i \right)^{q/2}\sum_{1 \le i_1 \le \ldots \ldots i_q \le N}
  j_{i_1 \ldots i_q } \psi_{i_1} \, \ldots \psi_{i_q} \ , \label{sykH}
\end{eqnarray}
where $q\leq N$ and $q= {\rm even}$.  The set of couplings $\left\{ j_{i_1
    \ldots i_q } \right\}$ are drawn from a random distribution, such
as a Gaussian one, described by 
\begin{eqnarray}
  \cP\left(  j_{i_1 \ldots i_q } \right) = {\rm exp} \left[ -
  \frac{N^3 j_{i_1 \ldots i_q }^2}{12 J^2}  \right] \ , \label{gaussian}
\end{eqnarray}
where $\cP$ denotes the probability distribution.  The gaussian
distribution for a random variable means the average value of the
couplings $j_{i_1 \ldots i_q }$ is zero and the two point average with
all indices contracted is non-vanishing,
\begin{eqnarray}
\left \langle j_{i_1 \ldots i_q } \right \rangle = 0 \ ,  \quad \left
  \langle j_{i_1 \ldots i_q }^2 \right \rangle = \frac{J^2 \left( q -1
  \right)!}{N^{q-1}} \ . 
\end{eqnarray}
The Majorana condition on the fermions simply means that they
satisfy the anti-commutation relation,
\begin{eqnarray}
\left\{ \psi_i , \psi_j \right\} = \delta_{ij} \ .
\end{eqnarray}
The Lagrangian corresponding to (\ref{sykH}) is given by
\begin{eqnarray}
&& S = \int d\tau L_{\rm E} \left ( \left\{\psi_i  \right\}, \left\{ 
\frac{d{\psi_i}}{d\tau}  \right\} \right) \ , \quad  L _{\rm E}= 
\frac{1}{2} \psi_i \frac{d\psi_i}{d\tau} - H \ , \label{sykL} \\
&& {\rm equivalently} \quad L = - \frac{1}{2} \psi_i
   \frac{d\psi_i}{dt} - H \ , \quad {\rm with} \quad t = - i \tau \ .
\end{eqnarray}
In the above $L_{\rm E}$ and $L$ corresponds to the Lagrangian in
Euclidean and Minkowski signatures, respectively.

\subsection{SYK model with Complex Fermions}
\label{sec:syk-model-cplx-ferm}

In order to introduce a chemical potential, we will explore the model
involving complex fermions.  This model has been studied earlier in
the condensed matter context \cite{Davison:2016ngz}, focussing on
transport properties and thermodynamics; and in the context of chaos
in \cite{Bulycheva:2017uqj}.  We are interested in the large $q$
expansion of the complex fermion model with an addition of a
non-vanishing chemical potential, which seems analogous to adding a
mass term.

The Hamiltonian for the SYK model with complex fermions is
\begin{equation}
H=\sum J_{i_1i_2...i_{q/2}i_{q/2+1}...i_q
}\psi^{\dagger}_{i_1}\psi^{\dagger}_{i_2}...
\psi^{\dagger}_{i_{q/2}}\psi_{i_{q/2+1}}....\psi_{i_q}  \ . \label{sykHC}
\end{equation}
In what follows we will use the notations and conventions used in
\cite{Davison:2016ngz}.  In addition to this interaction term we
introduce a chemical potential $\mu$.  We are interested in studying
the effect of a conserved charge on the chaotic behaviour of the
model.  Some of the earlier works \cite{Davison:2016ngz,
  Bulycheva:2017uqj} have analysed this model with either quartic
interactions or in the non-chaotic regime.  We will work in the large
$q$ limit and find out how the Lyapunov exponent changes as we tune in
the chemical potential.

\subsection{Free fermion propagator, with a chemical potential}

We define, following \cite{Davison:2016ngz}, the Green's function to
be: $G\left( \tau \right) = - \left \langle {\rm T} \psi\left(\tau
  \right) \psi^{\dagger} \left( 0\right) \right \rangle$, where the
symbol ${\rm T}$ stands for time-ordering and $\tau$ is the imaginary
time. The free fermion propagator, in the Fourier space, takes the
form: 
\begin{equation}
G(\mu,\omega) = \frac{1}{i\omega+\mu} \ ,
\end{equation}
which, in the real space, corresponds to the operator $ \left(
  -\partial_t +\mu
\right)$.  The two point function in the interacting theory, in the
large $q$ limit, can be expanded as:
\begin{equation}
G(\mu, \tau)=G_0(\mu, \tau)\left( 1+ \frac{g(\mu, \tau)}{q}+..\right)\ ,
\end{equation}
where $G_0(\mu, \tau)$ is the Fourier transform of the free
propagator, which at zero temperature it is given by,
\begin{equation}
G_0(\mu, \tau)=-e^{\mu \tau}\Theta(-\tau)  \ .
\end{equation}
Here
$\Theta$ is the Heaviside step function. At non-vanishing temperature,
however, it is obtained by evaluating the sum over Matsubara
frequencies that appear in the propagator,
$(i\omega_n+\mu)^{-1}$, which yields,
\begin{eqnarray}
G_0(\mu, \tau) & = & -\frac{e^{\mu\tau}}{e^{\mu\beta} + 1} \ ,   0 \leq \tau \leq \beta \ , \\
G_0(\mu,\tau) & = & \frac{e^{\mu\tau}}{e^{-\mu\beta} + 1} \ ,   -\beta \leq \tau \leq 0 \ .
\end{eqnarray}
The propagator for $\tau<0$ is obtained using the periodicity
$\tau\rightarrow \tau + \beta$.  The relative sign between $\tau<0$
and $\tau>0$ is a reflection of the fact that $G_0(\mu,\tau)$ is a
fermion propagator.  Finally, the function $g(\mu, \tau)$ is the
correction due to melonic diagrams to the free propagator, in the large $q$
limit.  In the next subsection we will derive a differential equation
for $g(\mu, \tau)$ and subsequently solve it.

\subsection{Differential equation for $g(\mu,\tau)$}

\begin{figure}[ht!]
\begin{center}
{\includegraphics[width=0.8\textwidth]{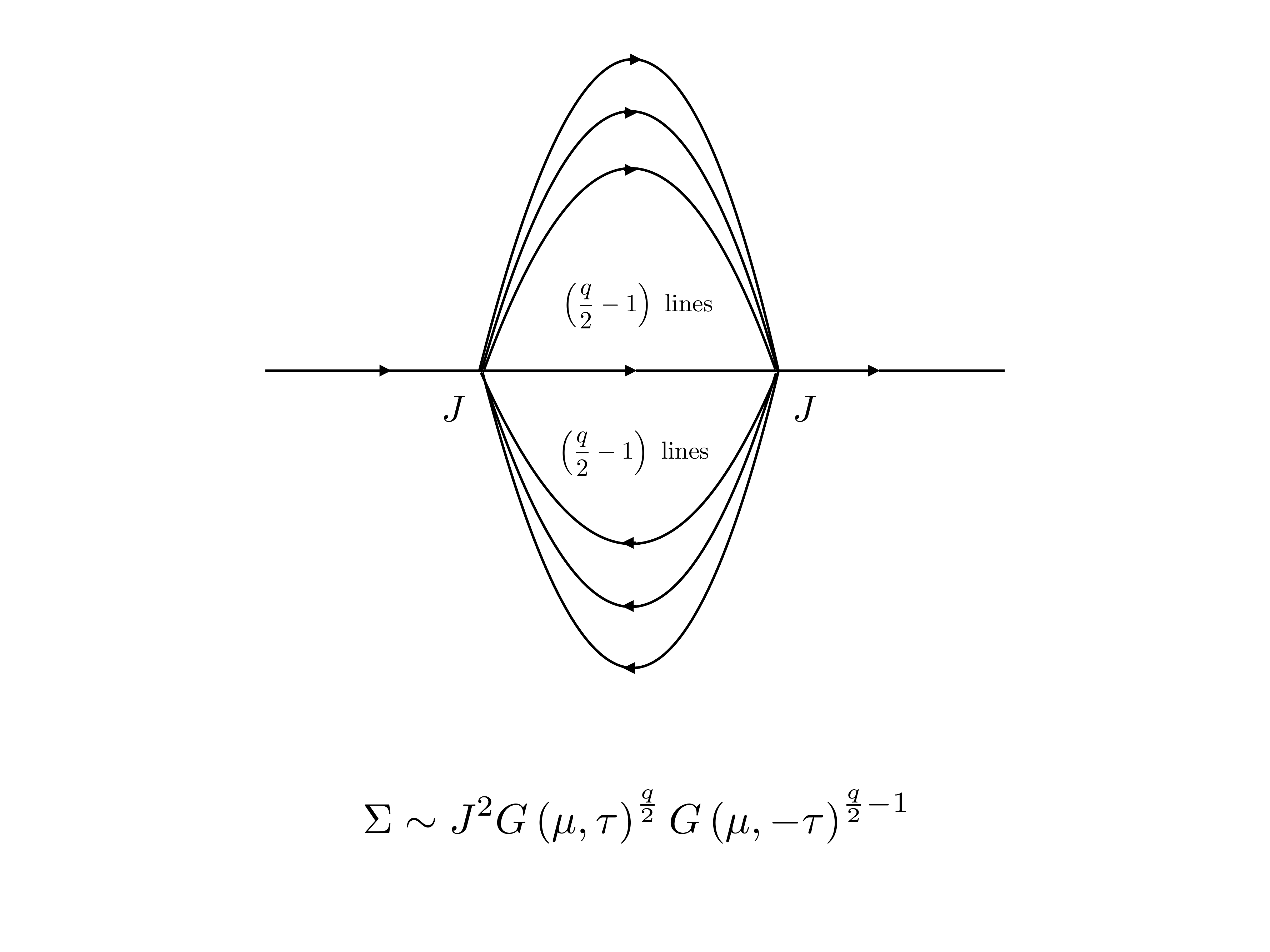}}
\caption{\small A diagrammatic representation of $\Sigma$. Each vertex
  is worth of strength $J$, and $\left( \frac{q}{2}-1 \right)$
  propagators run inside the loop in each direction. The direction of
  the arrows correlate with the sign of $\tau$ in the argument of the
  propagators. The overall direction of the diagram, from left to
  right, selects out two additional propagators running in this
  direction and hence the corresponding powers of $G$. } \label{sigma} 
\end{center}
\end{figure}

To derive the desired differential equation, we follow a simple
generalisation of the method discussed in
\cite{Maldacena:2016hyu}. First, note that, in the large $N$ limit,
all melonic Feynman diagram can be summed up to obtain the following
Schwinger-Dyson equation: 
\begin{eqnarray}
\frac{1}{G(\mu, \omega)} & = &   i \omega+\mu -\Sigma(\omega, \mu) \ , \\
\Sigma(\omega, \mu) & = & J^2(-1)^{q/2}(G(\mu, \tau))^{q/2}(G(\mu, -\tau))^{q/2-1}  \ . 
\end{eqnarray}
It is straightforward to derive the above Schwinger-Dyson equations by
summing up the one particle irreducible diagrams. Specifically, it is
straightforward to observe the second line above via Feynman
diagrammatic, see figure~\ref{sigma}. 

These Schwinger-Dyson equations take especially simple form in the $q
\to\infty$ limit.  In particular, the function $g(\mu, \tau)$ in this
limit appears in the exponential:
\begin{equation}
\label{eq:1}
\frac{1}{G(\mu, \omega)}=  i\omega+\mu -(i\omega+\mu)^2 \frac{f*g(\mu,\omega)}{2q} \ . 
\end{equation}
\begin{equation}\label{eq:2}
\Sigma(\mu,\tau)=\frac{J^2
  G_0(\mu,\tau)}{(2+2\cosh(\mu\beta))^{q/2-1}}e^{\frac{1}{2}(g(\mu,\tau)+g(\mu,-\tau))}
\ . 
\end{equation}
We can now identify the self energy contribution to the inverse
propagator as the Fourier transform of $\Sigma(\mu,\tau)$ appearing in
\eqref{eq:2}.  Taking the inverse Fourier transform of the self energy
contribution in \eqref{eq:1} we get the differential equation:
\begin{equation}
  \label{eq:3}
  (\partial_t -\mu)^2\left[ G_0(\mu,\tau)g(\mu,\tau)\right]=2\frac{q
  J^2
  G_0(\mu,\tau)}{2(2+2\cosh(\mu\beta))^{q/2-1}}e^{\frac{1}{2}(g(\mu,\tau)+g(\mu,-\tau))} \ .
\end{equation}
For $\tau > 0$ this equation reduces to:
\begin{equation}
  \label{eq:4}
  \partial_{\tau}^{2}g(\mu,\tau)=2\tilde{J}^2 e^{\frac{1}{2}(g(\mu,\tau)+g(\mu,-\tau))} \ ,
\end{equation}
where, 
\begin{eqnarray}
\tilde{J}^2 = \frac{q J^2}{2(2 + 2 \cosh(\mu\beta))^{\frac{q}{2}-1}} \ . 
\end{eqnarray}
It is worth pointing out at this point that this differential equation is
quite similar to that appearing in \cite{Maldacena:2016hyu}.  We will
solve this equation analytically in the next section.  

Before moving further, a few comments regarding the large $q$
result are in order.  It is straightforward to check that, if one goes beyond the
leading order in $(1/q)$-expansion, the Schwinger-Dyson equation again
rearranges itself to the differential equation of the type discussed
above, with the same effective coupling $\tilde{J}$.   

To see this explicitly let
us first notice that the $\mu$ dependence of $\tilde{J}$ comes only from the
free part. If we look at the behavior of the self-energy contribution
at $O(\frac{1}{q^2})$ we find for $\mu=0$ case, the terms take the form
\begin{equation}
  \label{eq:6}
  J^2\left(1+\frac{g(\tau)}{q}+\frac{g^{\prime}(\tau)}{q^2}+...\right)^{q-1}\ .
\end{equation}
The equation for function $g^{\prime}$ cannot be obtained by simply
exponentiating it, as was done for the leading correction, namely
$g(\tau)$.  We instead have an asymptotic series expansion in
$\frac{1}{q}$. Now if we turn on finite $\mu$ then from
the self-energy expression we get,
\begin{equation}
  \label{eq:5}
  \frac{J^2}{2(2+2\cosh(\mu\beta))^{q/2-1}}\left(1+
  \frac{g(\mu,\tau)}{q}+\frac{g^{\prime}(\mu,\tau)}{q^2}+.. 
  \right)^{\frac{q}{2}}\left(1+
  \frac{g(\mu,-\tau)}{q}+\frac{g^{\prime}(\mu,-\tau)}{q^2}+..
  \right)^{\frac{q}{2}-1}
\end{equation}
The form is exactly like in the SYK model.  As a result the equation that we
would obtain in this case will be identical to that for $g^{\prime}$
in the SYK model.  In other words even for finite $\mu$, the effective
coupling constant $\tilde{J}$ remains unaltered even at higher order
in $1/q$.  The emergence of one effective coupling is an inherent
feature of this asymptotic expansion in $(1/q)$.

\section{Calculating the retarded kernel}

The right hand side of the differential equation \eqref{eq:4} is
symmetric under $\tau \to - \tau$, whereas on the left hand side we
switch from $g(\mu, \tau)\to g(\mu, - \tau)$.  We can therefore send
$\tau \to - \tau$, and subsequently obtain the resulting equation for
$g(\mu, - \tau)$. The solutions to the differential equations are
exactly of the Maldacena-Stanford form \cite{Maldacena:2016hyu}, and
are given by
\begin{eqnarray}
e^{g \left(\mu, \pm \tau \right)} =
  \frac{\cos^2\left(\frac{\pi\nu}{2}\right)}{\cos^2\left(\pi\nu
  \left(\frac{\tau}{\beta} \mp \frac{1}{2}\right) \right)} \ , \quad
  {\rm with} \quad  \beta\tilde{J} = \frac{\pi\nu}{\cos \left(
  \frac{\pi\nu}{2} \right)} \ . \label{solg} 
\end{eqnarray}
Note that, the parameter $\nu$ that naturally emerges here contains
information about the two independent UV-couplings: $\beta J$ and
$\beta\mu$.

\subsection{The retarded Green's function}

We begin by defining the retarded Green's function
\begin{equation}
G_{\rm R} \left(\mu, t \right) = \lim_{\epsilon\rightarrow 0+} \left[
  G_ {>} \left(\mu, it + \epsilon \right) - G_ {<} \left(\mu, it -
    \epsilon \right)\right] \Theta(t) \ . 
\end{equation}
In the $q\rightarrow\infty$ limit, we obtain:
\begin{equation}
G_{\rm R} \left(\mu, t \right) = -e^{i\mu t}\Theta(t) \ .
\end{equation}
The above result, in the limit $\mu \rightarrow 0$, yields: $G_{\rm
  R}(t) = \Theta(t)$ which is the expected answer.  We can also
define: 
\begin{equation}
G_{\rm R} \left(\mu, -t \right) = \lim_{\epsilon\rightarrow 0+}
\left[G_ {>} \left(\mu,- \left(it + \epsilon \right) \right) - G_ {<}
  \left(\mu, - \left(it - \epsilon \right) \right)\right] \Theta(t) \
, 
\end{equation}
which implies $G_{\rm R} \left(\mu,-t \right) = e^{-i\mu
  t}\Theta(t)$.

\subsection{The retarded kernel}

\begin{figure}[ht!]
\begin{center}
{\includegraphics[width=0.8\textwidth]{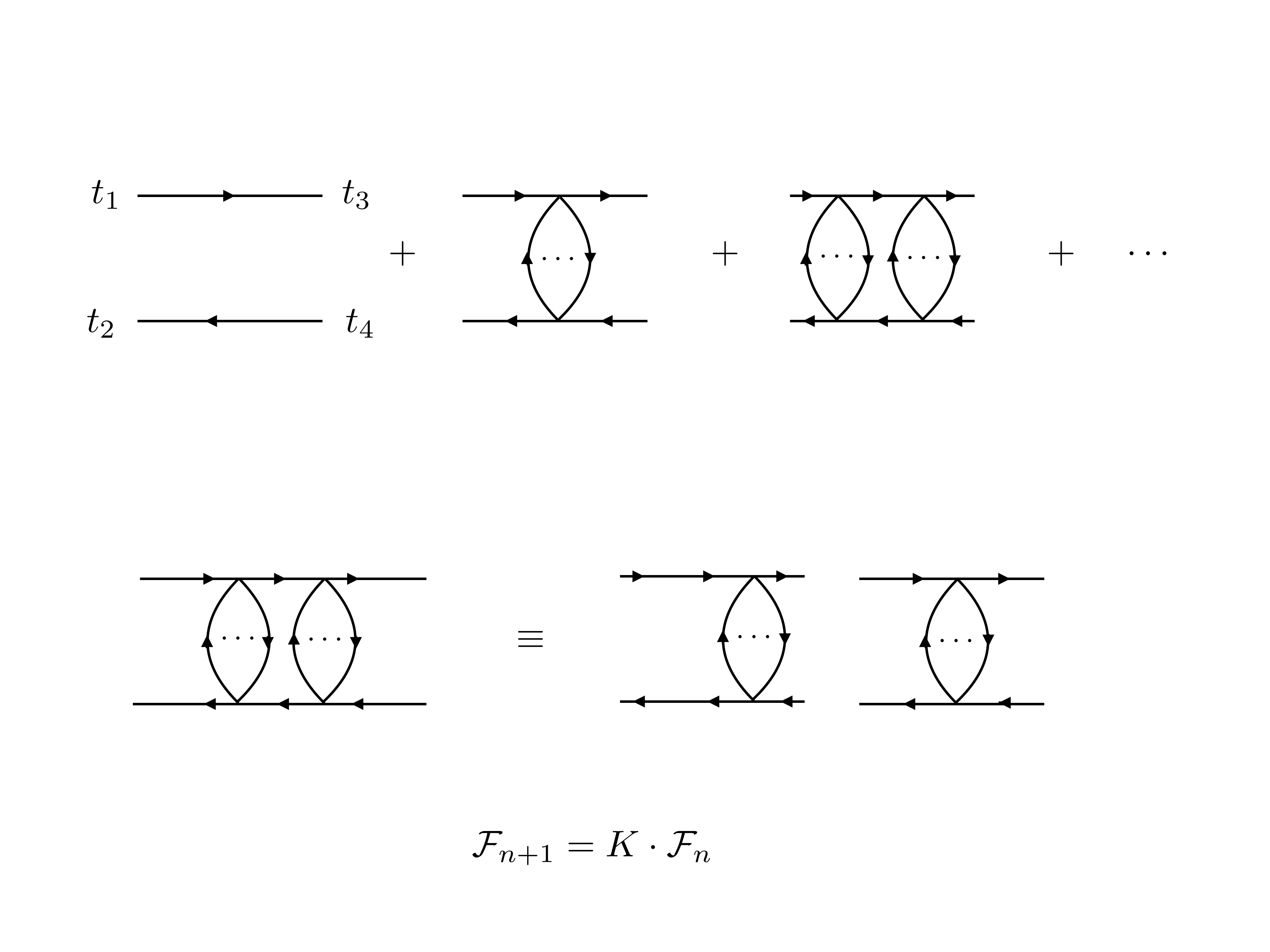}}
\caption{\small A diagrammatic representation of the four point
  function calculation, in the large $N$ limit. First, only the ladder
  diagrams contribute, as shown in the first row here. Second, from
  the structure of the diagrams, one obtains an iterative process to
  generate ${\cal F}_{n+1}$ from ${\cal F}_n$, composing with a
  kernel.} \label{kernelK} 
\end{center}
\end{figure}

Now we analyze the four-point function. In the large $N$ limit, the
four-point function can be expanded in a series of $(1/N)$ and, here,
we will only compute the the leading $(1/N)$-contribution, in which
only the ladder diagrams contribute. Since we are working with complex
fermions, the only non-trivial four-point function is given by 
\begin{eqnarray}
\frac{1}{N^2} \sum_{i,j=1}^N \left\langle T \left(
  \psi_i(t_1)\psi_i^{\dagger}(t_2)\psi_j^{\dagger}(t_3)\psi_j(t_4)
  \right)  \right\rangle = G \left(t_{12}\right) G \left(t_{34}\right)
  + \frac{1}{N} {\cal F} \left( t_1, t_2, t_3, t_4 \right) + \ldots
  \nonumber\\ 
\end{eqnarray}
The contribution at order $(1/N)$ is collectively denoted by ${\cal F}
= \sum_n {\cal F}_n$, where $n$ is the number of rungs in the
corresponding ladder diagram. We refer to \cite{Maldacena:2016hyu} for
more details. The composition rule is pictorially represented in
figure~\ref{kernelK}. 

At large $N$, the summation over the ladder diagrams can be performed
by expressing ${\cal F}_{n+1}$ in terms of ${\cal F}_n$ integrated,
weighted with a kernel, as also pictorially shown in
figure~\ref{kernelK}: 
\begin{eqnarray}
{\cal F}_{n+1} \left( t_1, t_2, t_3, t_4 \right) = \int dt dt' K_{\rm
  R} \left( t_1, t_2 ; t, t' \right) {\cal F}_{n} \left(t, t', t_3,
  t_4 \right) \ , 
\end{eqnarray}
where the kernel, denoted above by $K_{\rm R}$, is given by
\begin{eqnarray}
  K_{\rm R} \left(t_1,t_2,t_3,t_4 \right) = (-1)^{q/2}J^2(q-1) 
  &&G_{\rm R}\left(\mu, t_{13}\right) G_{\rm R} \left(\mu, -t_{24}\right)
     \nonumber\\ 
  && \left[G_{\rm lr}(\mu, t_{34})\right]^{q/2-1}\left[G_{\rm
     lr}(\mu,-t_{34})\right]^{q/2-1} \ . 
\end{eqnarray}
Here $G_{\rm lr} \left(\mu, t \right)$ is the Wightman function, which
is essentially given by the propagator evaluated at complex time, and
in the large $q$ limit we get:
%
\begin{equation}
 \left[G_{\rm lr}(t)\right]^{q/2-1} \left[G_{\rm lr}(-t)
 \right]^{q/2-1} =  \left[G(it+\beta/2) \right]^{q/2-1}
 \left[G(-it+\beta/2) \right]^{q/2-1} \ . 
\end{equation}
%
The above is consistent with
interpreting the propagator $G(\mu,-t)$ as the fermion moving backward
in time, or the anti-fermion moving forward in time. This is why a
separation along the thermal circle picks up a relative sign.

Finally, we obtain:
\begin{equation}
(-1)^{q/2}J^2(q-1) \left[G_{\rm lr}(t) \right]^{q/2-1} \left[G_{\rm lr}(-t) \right]^{q/2-1}=(-1)^{q-1}\frac{2\pi^2\nu^2}{\beta^2 \cosh^2\left(\frac{\pi\nu t}{\beta}\right)} \ .
\end{equation}
Using this, the complete retarded kernel is given by
\begin{eqnarray}
K_{\rm R} \left(t_1,t_2,t_3,t_4 \right) & = & -(-1)^{q-1}e^{i\mu \left(t_{13}-t_{24} \right)} \frac{2\pi^2\nu^2\Theta\left(t_{13} \right) \Theta \left(t_{24}\right)}{\beta^2 \cosh^2 \left(\frac{\pi\nu t_{34}}{\beta}\right)} \label{kernel} \\
& = & e^{i\mu \left(t_{12}-t_{34} \right)} \frac{2\pi^2\nu^2 \Theta \left(t_{13} \right) \Theta \left(t_{24} \right)}{\beta^2 \cosh^2 \left(\frac{\pi\nu t_{34}}{\beta}\right)} \ .
\end{eqnarray}
The last equality follows from the fact that $q$ is even.

\section{Exploring the chaos regime}

\begin{figure}[ht!]
\begin{center}
{\includegraphics[width=0.6\textwidth]{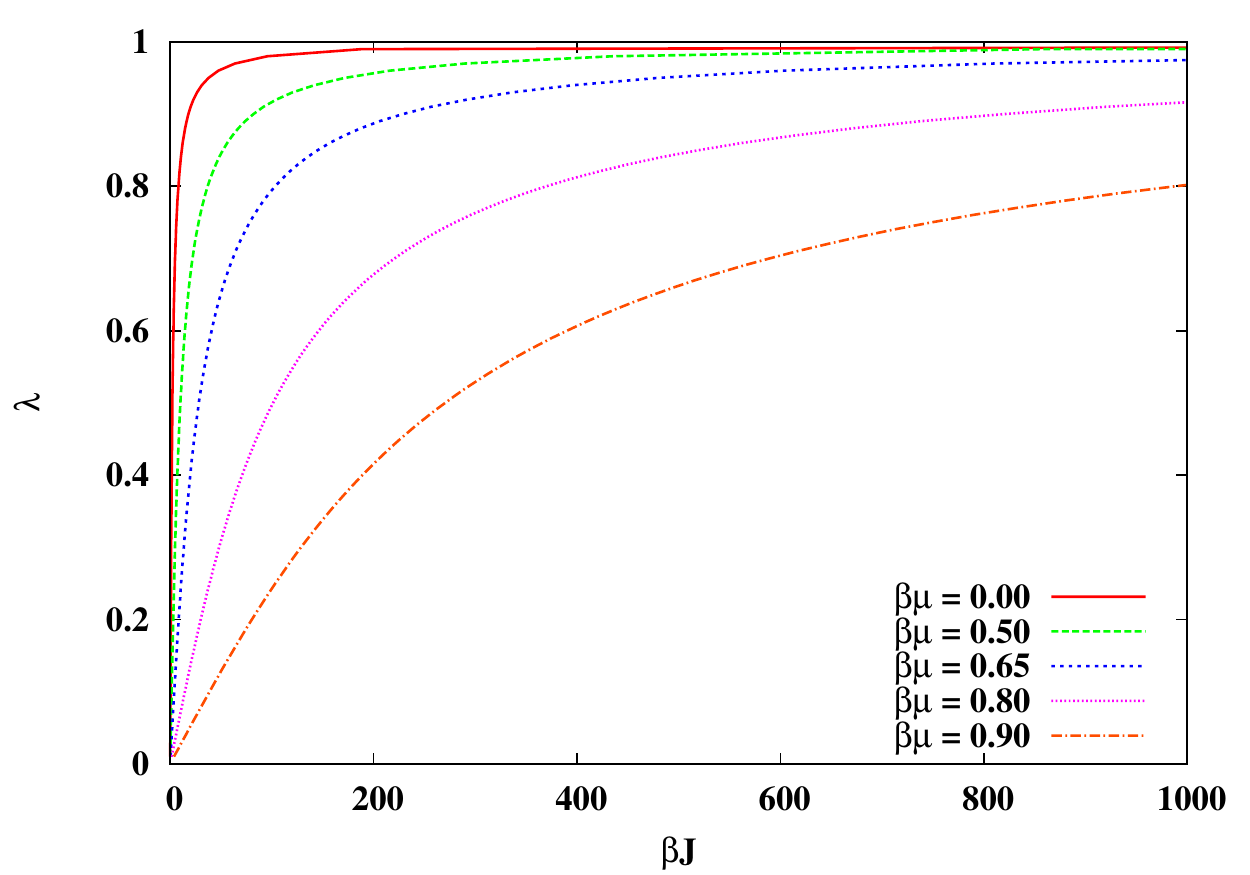}}
\caption{\small The Lyapunov exponent $\lambda$ is normalised and
  takes values between 0 and 1. This figure shows dependence of
  $\lambda$ on $\beta J$ for different values of $\beta\mu$} \label{LambdaJ} 
\end{center}
\end{figure}

\begin{figure}[ht!]
\begin{center}
{\includegraphics[width=0.6\textwidth]{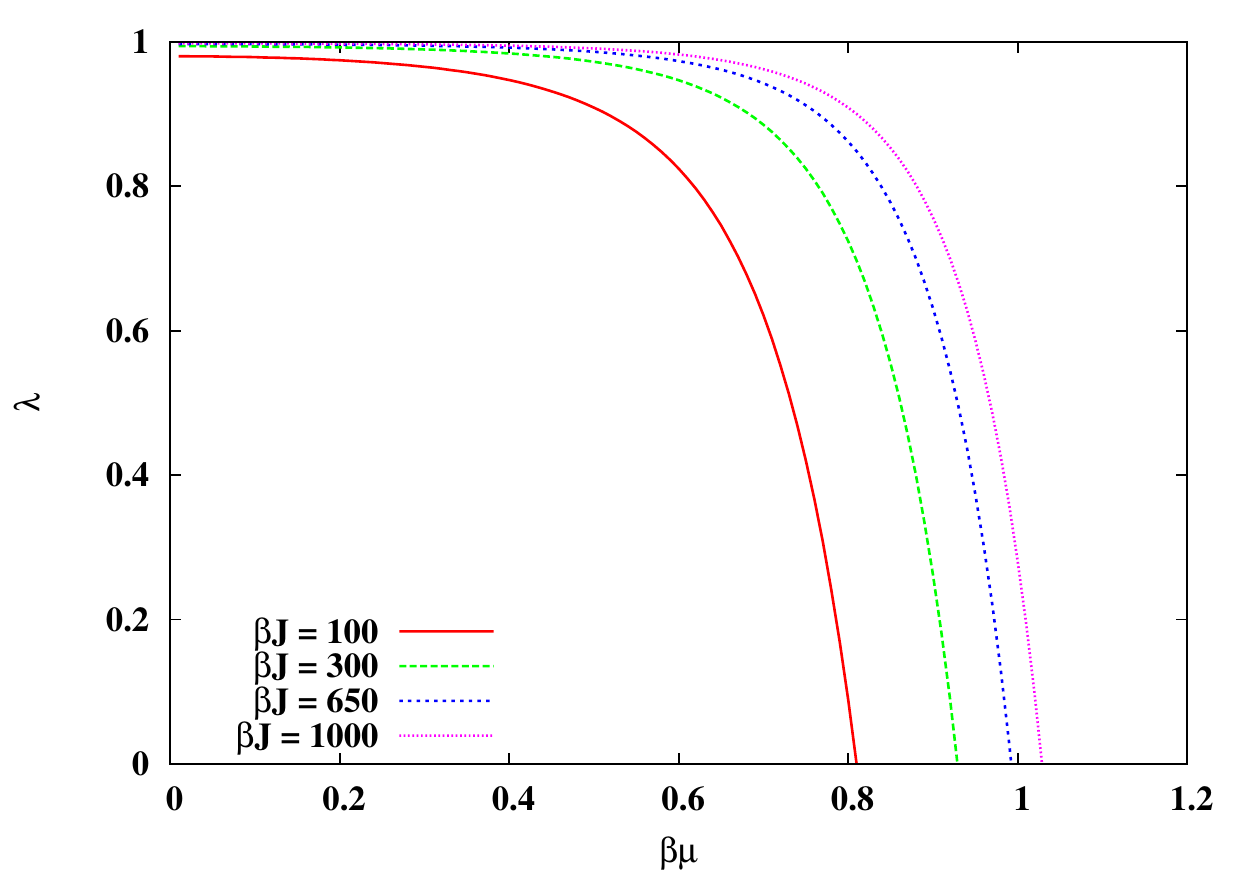}}
\caption{\small The Lyapunov exponent $\lambda$ is again normalised and
  takes values between 0 and 1. This figure shows dependence of
  $\lambda$ on $\beta\mu$ for different values of $\beta J$} \label{Lambdamu} 
\end{center}
\end{figure}

So far, we have obtained the retarded kernel for four fermion fields
placed at four arbitrary points on the thermal circle, denoted
respectively by $t_1, \ldots, t_4$. To extract the chaos behaviour,
one needs to calculate the OTO correlation in real time, separating
the fermions by a quarter of the thermal circle\cite{Kitaev}. We want to
compute the following OTO correlation: 
\begin{eqnarray}
{\cal F} \left( t_1, t_2 \right) = {\rm Tr} \left[ y \psi_i(t_1) y \psi_i^{\dagger}(0) y \psi_j^{\dagger}(t_2) y \psi_j(0)  \right] \ , \quad y = \rho(\beta)^{1/4} \ .
\end{eqnarray}
In the limit $t_1, t_2 \to \infty$, the diagram with zero rung is
suppressed and thus ${\cal F}(t_1, t_2)$ is an eigenfunction of the
retarded kernel $K_{\rm R}$, with an eigenvalue one. This statement
translates into an integral equation of the following form: 
\begin{eqnarray}
{\cal F} \left(t_1,t_2 \right) & = &  \int_{-\infty}^{\infty} \int_{-\infty}^{\infty} dt_3 dt_4 K_{\rm R} \left(t_1,t_2,t_3,t_4\right) {\cal F}\left(t_3,t_4 \right) \\
& = & \int_{-\infty}^{\infty}\int_{-\infty}^{\infty} dt_3 dt_4 e^{i\mu \left(t_{12}-t_{34} \right)} \frac{2\pi^2\nu^2 \Theta \left(t_{13}\right) \Theta \left(t_{24}\right)}{\beta^2 \cosh^2 \left(\frac{\pi\nu t_{34}}{\beta}\right)}{\cal F} \left(t_3,t_4 \right) \ .
\end{eqnarray}
Choosing an exponential-ansatz for ${\cal F}\left(t_3, t_4 \right)$ of the form
\begin{eqnarray}
{\cal F} \left(t_3,t_4 \right) = e^{\frac{\pi\nu}{\beta} \left(t_3 + t_4 \right)}\frac{e^{i\mu t_{34}}}{\cosh\left(\frac{\pi\nu t_{34}}{\beta}\right)} \ ,
\end{eqnarray}
yields:
\begin{eqnarray}
{\cal F}\left(t_1,t_2 \right) & = &  e^{i\mu t_{12}}\int_{-\infty}^{t_1}\int_{-\infty}^{t_2}
dt_3 dt_4 \frac{2\pi^2\nu^2e^{\frac{\pi\nu}{\beta}
\left(t_3 + t_4 \right)}}{\beta^2 \cosh^3 \left(\frac{\pi\nu
t_{34}}{\beta}\right)} \\ 
& = & e^{\frac{\pi\nu}{\beta} \left(t_1 + t_2 \right)}\frac{e^{i\mu
      \left(t_{12} \right)}}{\cosh\left(\frac{\pi\nu
      t_{12}}{\beta}\right)} \ . 
\end{eqnarray}
This implies, following the subsequent steps outlined in
\cite{Maldacena:2016hyu}, that the Lyapunov exponent is given by 
\begin{equation}
\lambda_{\rm L} = \frac{2\pi}{\beta}\nu \ ,
\end{equation}
where $\nu$ is given in equation (\ref{solg}). In the two extreme
limits, we easily get: 
\begin{eqnarray}
\lambda_{\rm L} & = & \left( 2 \tilde{J} \right) + \ldots \ , \quad
                      {\rm as} \quad \nu \to 0 \quad \iff  \quad \beta
                      \tilde{J} \to 0 \ , \\ 
& = & \frac{2\pi}{\beta} \left( 1 - \frac{2}{\beta \tilde{J}} \right)
      \ , \quad {\rm as} \quad \nu \to 1 \quad \iff  \quad \beta
      \tilde{J} \to \infty \ . 
\end{eqnarray}
In terms of the IR emergent coupling $\beta\tilde{J}$, the dependence
is identical to the one observed in \cite{Maldacena:2016hyu}, however,
in terms of the original parameters $\{\beta J, \beta \mu \}$ defining
the system, there is a non-trivial dependence of the Lyapunov
exponent.  The figure \ref{LambdaJ}, shows behaviour of
$\lambda=\beta\lambda_L/2\pi$, which is the normalised Lyapunov
exponent, as a function of the coupling $\beta J$ for various values
of $\beta\mu$.  Similarly the figure \ref{Lambdamu} shows variation of
$\lambda$ as a function of $\beta\mu$ for different values of $\beta J$.

Before concluding this section, let us make some comments regarding tuning the chaotic properties of SYK-type models. In \cite{Garcia-Garcia:2017bkg}, a two-body infinite-range random interaction between Majorana fermions was introduced, in addition to the four-fermi interaction in the SYK model. It was found that this interaction can tune the Lyapunov exponent down, and in fact, push it all the way to zero, similar to what we have observed above. However, the precise dependence of the Lyapunov exponent with the one-body interaction strength is different compared to our results. 

The Hamiltonian considered in \cite{Garcia-Garcia:2017bkg} is of the following form:
\begin{eqnarray}
H = \sum_{1 \le i_1\le i_2\le i_3\le i_4 \le N} J_{i_1 i_2 i_3 i_4} \ \psi_{i_1} \psi_{i_2} \psi_{i_3} \psi_{i_4} + i \sum_{1 \le i_1 \le  i_2 \le N} k_{i_1 i_2 } \ \psi_{i_1} \psi_{i_2} \ ,
\end{eqnarray}
where $J_{i_1 i_2 i_3 i_4}$ are chosen from a familiar Gaussian ensemble, and the couplings $k_{i_1 i_2}$ denote the infinite-range interaction and $\psi_i$'s are Majorana fermions. Assuming $N$ is even, we can consider a particularly special case, in which $k_{i_1 i_2}$ are non-random, and are characterized by a particularly nearest neighbour interaction:
\begin{eqnarray}
k_{i_1 i_2} & = &  k \delta_{i_1+1, i_2 } \quad {\rm if} \quad i_1 = {\rm odd} \ , \nonumber\\
& = & 0 \quad {\rm otherwise} \ . \label{kijnear}
\end{eqnarray}
The interaction term is now particularly simple: 
\begin{eqnarray}
&& H_{\rm int} = i \sum_{i={\rm odd}}^N k_{i, i+1} \ \psi_i \psi_{i+1} \equiv  \Psi^{\dagger} \cK \Psi \ , \\
&& {\rm where} \quad \Psi^{\dagger} = \left( \psi_1, \psi_2 , \ldots \psi_N \right) \ . 
\end{eqnarray}
Evidently, the $^{\dagger}$ operation is equivalent to the transpose operation since we are dealing with Majorana fermions. The matrix $\cK$ contains the information about the nearest-neighbour interaction of (\ref{kijnear}). It is easy to diagonalize the coupling matrix $\cK$, and the resulting eigenvalues are: $\left( \frac{N}{2} \right)$ copies of $\left( + \frac{k}{2} \right)$ and $\left( \frac{N}{2} \right)$ copies of $ -  \left( \frac{k}{2} \right)$. Suppose that $\chi_{a}^{+}$, with $a = 1 , \ldots , N/2$, eigenvectors have positive eigenvalues and $\chi_{a}^{-}$, with $a = 1 , \ldots , N/2$, eigenvectors have negative eigenvalues. It is also straightforward to check that: $\left(\chi^{+}\right)^{\dagger} = \chi^{-}$, thus we can drop the superscript, and subsequently the interaction term can be written as:
\begin{eqnarray}
H_{\rm int} = k \sum_{a}^{N/2} \chi_a^\dagger \chi_a  \ , \quad {\rm where} \quad \left\{\chi_a^\dagger, \chi_b \right\} = 2 \delta_{ab} \ .
\end{eqnarray}
We can now rewrite the four-body interaction in the complex $\chi$-basis. Since our starting point did not preserve the U$(1)$-symmetry of the complex fermion model in (\ref{sykHC}), the full resulting Hamiltonian does not match with the complex fermion model with $q=4$. However, in the UV, with $(J / k) \to 0$, the four-point interaction is negligible and the two systems are physically equivalent. In the IR, the two systems are completely distinct.

\section{Flavoured Complex fermions with a chemical potential}

Let us now generalise this set up, where instead of a U$(1)$ symmetry
we have $N_f$ number of flavoured fermions with a global SU$(N_f)$
flavour symmetry, similar to the model considered in \cite{Gross:2016kjj}.  The fermions now carry two indices,
$\Psi^{\alpha}_i$. Here the $\alpha$ is the flavour index where as $i$
is the site index. One has the following operator algebra:
\begin{equation}
\lbrace \Psi_i^{\alpha},\Psi_j^{\beta}\rbrace=\lbrace \Psi_i^{\alpha
  \dagger},\Psi_j^{\beta \dagger}\rbrace=0  \ , \quad     \lbrace
\Psi_i^{\alpha},\Psi_j^{\beta \dagger}\rbrace=\delta_{ij}\delta^{\alpha\beta} \ . 
\end{equation}
It is a trivial matter to find first the kinetic term without
introducing the chemical potential $\mu$ it given by
\begin{equation}
-\int d\tau\Psi^{\alpha\dagger}_i\partial_{\tau}\Psi^{\alpha}_i \ . 
\end{equation}
Here repeated indices are summed over unless stated otherwise.

The SU$(N_f)$ invariant two point function in this case will be given by
\begin{equation}
G(\tau)=\langle\Psi^{\alpha}_i(\tau)\Psi^{\alpha\dagger}_j(0)\rangle\equiv\frac{N_f {\rm sgn}(\tau)}{2}\delta_{ij} \ . 
\end{equation} 
If we absorb this factor of $N_f$ into the overall normalization of
the kinetic piece then we observe that now if one introduces a
conserved charge $\mu$ then the relevant operator is:
$$\frac{\mu}{N_f}\Psi^{\alpha}_i\Psi^{\alpha\dagger}_i.$$

We know that the interaction term should be a gauge singlet. We also
require that, upon imposing reality condition on the fermions,
this interaction should reduce to the corresponding interaction
term in the Gross-Rosenhaus model. Under this, we intuitively write down
the interaction term as:
\begin{equation}
\frac{1}{N_f^{q/2}}J_{i_1....i_q}\Psi^{\alpha_1\dagger}_{i_1}....
\Psi^{\alpha_{q/2}\dagger}_{i_{q/2}}\Psi^{\alpha_{q/2}}_{i_{q/2+1}}...\Psi^{\alpha_1}_{i_q} \ . 
\end{equation}
Now we just use the melon diagrams to figure out the 1PI effective
self energy contribution. Essentially, as before, we observe that from the
diagramatics one obtains:
$$\Sigma(\tau) = \frac{C^{N_f}_{\frac{q}{2}}}{N_f^q}J^2 \left[G(\tau)
\right]^{q/2} \left[G(-\tau)\right]^{q/2-1}. $$ 
So, one can redefine the coupling strength as:
$J_{\rm eff}^2 = \frac{C^{N_f}_{\frac{q}{2}}}{N_f^q}J^2$. This means that,
if we have multiple groups of flavours, then the relative strength of
the effective couplings scale according to the above relation.  Hence,
again we get back the same set of Schwinger-Dyson equations which we
have already solved.

We already see the emergence of an effective coupling:
\begin{eqnarray}
J_{\rm eff}^2 = \frac{1}{N_f^q} \frac{N_f!}{\left(\frac{q}{2}\right)! \left( N_f - \frac{q}{2}  \right) !} J^2 \ ,
\end{eqnarray}
which, in the limit $q \gg1$, $N_f \gg 1$ such that $N_f \gg q$, naively, yields:
\begin{eqnarray}
J_{\rm eff}^2 = \frac{1}{N_f^q} \frac{1}{\left(\frac{q}{2}\right)!} J^2 \to 0 \ .
\end{eqnarray}
Thus, with a very large global symmetry, the emergent coupling is very
weak. This implies that the resulting chaotic behaviour will be
accompanied with a vanishingly small value of the Lyapunov
exponent. Thus, we can tune the chaotic behaviour with a global
flavour symmetry, as well. 

\section{Conclusion}
\label{sec:conclusion}

In this article, we have explored and demonstrated a tuneable Lyapunov
exponent by introducing conserved charges in the system, even when the
charge is a simple U$(1)$. We have considered SYK-type models, with
complex fermions and a $q$-body all-to-all randomized interaction, in
the $q \to \infty$ limit. For these models, we have explicitly
demonstrated that a non-vanishing chemical potential has an
exponentially large dominance over the $q$-body interaction coupling
strength, in determining the chaos behaviour. It is expected, from the
structure of the Schwinger-Dyson equations, that similar features hold
for the tensor models\cite{Witten:2016iux}, which share many
interesting properties of the SYK-type interaction, but without the
disorder averaging.

There are various interesting directions for future
explorations. Given the results above, one may explore higher
dimensional generalizations of the SYK-model, {\it e.g.}~the model in
\cite{Turiaci:2017zwd}, with an introduction of conserved charges. One
would, na\'{i}vely, expect a similar behaviour of the resulting
Lyapunov exponent for the higher dimensional models; however, it would
be very interesting to check how the details fall into the right
places. Staying within the theme of a tuneable chaos, motivated by the
similarities of SYK-model behaviour and random matrix behaviour at
late times, it is natural to incorporate the effect of conserved
charges in random matrix theories and analyze the consequences at late
times\cite{Garcia-Garcia:2016mno,Cotler:2016fpe,Garcia-Garcia:2017pzl}.

From a holographic perspective, our analysis suggests that by
introducing bulk gauge fields that correspond to introducing chemical
potentials for the dual boundary theory, one should be able to do away
with chaos completely, or, at least, should be able to tune down the
Lyapunov exponent from its' maximal value. This would be an
interesting aspect to check explicitly. Towards that, one presumably
begins with a gravity description in {\it e.g.}~$(d+1)$-dimensional
bulk with AdS-asymptotic, and studies a scattering problem, {\it a la}
\cite{Shenker:2014cwa}, in the presence of a global charge. On a
similar note, it is also very intriguing to explore the possibility of
constructing an SYK-type model from explicit D-brane construction in
string theory, with or without global charges. One natural obstacle,
for the SYK-type interaction, is to realize the dynamical origin of
disorder averaging from the brane picture. Perhaps the large $N$
tensor models can emerge more naturally in such scenarios. We are
currently exploring some of these issues further.

\noindent {\bf Acknowledgments:} We thank Ashoke Sen and K. Sengupta
for many interesting discussions.  RB and SC would like to thank Saha
Institute for hospitality and DPJ would like to thank IMSc for
hospitality.  Work of SC was partly supported by the Infosys
scholarship for senior students.



\end{document}